\documentclass[aps,prl,twocolumn,showpacs,preprintnumbers,amsmath,amssymb,floatfix,superscriptaddress]{revtex4-1}
\usepackage{graphicx}
\usepackage{amsfonts}
\usepackage{CJK}
\usepackage{bm}
\usepackage{color}
\usepackage{xcolor}
\usepackage{array}
\usepackage[T1]{fontenc}
\usepackage[colorlinks = true,linkcolor = blue,urlcolor = blue,citecolor = blue,anchorcolor = blue]{hyperref}

\begin{document}
\bibliographystyle{apsrev4-1}

\title{Geometry protected probabilistic structure in many-body dynamics}
\author{Yue Liu\textsuperscript{1,3}}
\author{Chushun Tian\textsuperscript{2}}\email{ct@itp.ac.cn}
\author{Dahai He\textsuperscript{1}}\email{dhe@xmu.edu.cn}

\noaffiliation
\affiliation{Department of Physics and Jiujiang Research Institute, Xiamen University, Xiamen 361005, Fujian, China }
\affiliation{Institute of Theoretical Physics, Chinese Academy of Sciences, Beijing 100190, China }
\affiliation{Yukawa Institute for Theoretical Physics, Kyoto University, Kitashirakawa Oiwakecho, Sakyo-ku, Kyoto 606-8502, Japan}

\date{\today}
\begin{abstract}
  Insomuch as statistical mechanics circumvents the formidable task of addressing many-body dynamics, it remains a challenge to derive macroscopic properties from a solution to Hamiltonian equations for microscopic motion of an isolated system. Launching new attacks on this long-standing problem --- part of Hilbert's sixth problem --- is urgently important, for focus of statistical phenomena is shifting from a fictitious ensemble to an individual member, i.e. a mechanically isolated system. Here we uncover a common probabilistic structure, the concentration of measure, in Hamiltonian dynamics of two families of systems, the Fermi-Pasta-Ulam-Tsingou (FPUT) model which is finite-dimensional and (almost) ergodic, and the Gross-Pitaevskii equation (GPE) which is infinite-dimensional and suffers strong ergodicity breaking. That structure is protected by the geometry of phase space and immune to ergodicity breaking, leading to counterintuitive phenomena. Notably, an isolated FPUT behaves as a thermal ideal gas even for strong modal interaction, with the thermalization time analogous to the Ehrenfest time in quantum chaos, while an isolated GPE system, without any quantum inputs, escapes the celebrated ultraviolet catastrophe through nonlinear wave localization in the mode space, and the Rayleigh-Jeans equilibrium sets in the localization volume. Our findings may have applications in nonlinear optics and cold-atom dynamics.
\end{abstract}
\maketitle
  Statistical mechanics turns a classical mechanical description of individual many-body system into a probabilistic description of a fictitious ensemble of identical individuals. The mathematical foundations of this treatment, that partially motivated Hilbert's sixth problem \cite{Gorban18}, remain elusive. There are two widely accepted physical interpretations. One attributes the probabilistic description to that a system is hardly isolated in practice, but coupled to stochastic environments such as a heat bath \cite{Landau37}. The other resorts to the (classical) ergodic hypothesis, i.e. the infinite time average of a dynamical quantity equals the microcanonical ensemble average \cite{Benettin94}. Recent progresses in many-body physics have deepened questioning of these interpretations. First, realizing isolated systems is now well within the experimental reach, and breakthroughs have been made for statistical mechanics of isolated systems albeit quantum \cite{Deutsch91,Srednicki94,Rigol08,Popescu06,Lebowitz06,Kaufman16}. Secondly, the ergodic hypothesis provides no information on the temporal evolution of statistical properties. Nor does it apply to systems carrying infinite number of degrees of freedom, e.g. field-matter interacting systems and wave turbulence whose statistical behaviors have attracted much interest \cite{Wang22,Zakharov12}. Thus a major step forward is to explore novel probabilistic structures, formed during microscopic Hamiltonian dynamics and insensitive to ergodic properties, and ensuing macroscopic phenomena.

  State-of-art numerical experiments have been launched for various isolated classical many-body systems to study statistical properties of a single phase trajectory, that records the history of system's microscopic state \cite{Wang22,Raedt13,Flach17,Flach19,Lebowitz22}. Results were compared to traditional theories based on the fictitious ensemble, and display important anomalies among which are the following; see Supplementary Information (SI) \cite{SI} for more discussions. In some integrable systems the Gibbs distribution for a subsystem results directly from a phase trajectory \cite{Raedt13}, showing a parallel to subsystem thermalization of an isolated quantum system \cite{Popescu06,Lebowitz06,Yang15}. More surprisingly, a classical field-matter interacting model, though carrying infinite number of degrees of freedom, was found to escape the ultraviolet catastrophe by self-generating a high-frequency cutoff of the blackbody spectrum, and statistics of fluctuations along the phase trajectory is in concert with the Stefan-Boltzmann law \cite{Wang22}. This disrupts the common belief for over one century, i.e. the paradox of ultraviolet catastrophe cannot be solved in classical physics. These findings, though still lack of theoretical explanations, indicate rich classical statistical phenomena displayed by isolated many-body system, which are beyond that entailed by ergodicity.

  Here we develop a mathematical theory for two families of classical many-body systems carrying opposite ergodic properties, the FPUT model and the GPE, and show that their phase space is endowed a common probabilistic structure, the concentration of high-dimensional Gaussian measure, by a phase trajectory. This structure is protected by the phase-space geometry, immune to ergodicity breaking. The FPUT model is an anharmonic chain, whose Hamiltonian dynamics takes place in a finite-dimensional phase space and is (almost) ergodic \cite{Fermi55}. Historically, studies of its statistical behaviors led to the birth of several fields \cite{Izrailev05,Gallavotti08}. The GPE describes dynamics of Bose-Einstein condensates \cite{Pitaevskii03} and propagation of classical waves in nonlinear media \cite{Zakharov12}. Hamiltonian dynamics entailed by GPE takes place in an infinite-dimensional phase space and suffers strong ergodicity breaking. Thermalization processes of GPE systems have received many experimental attentions recently \cite{Christodoulides22}.

  The uncovered probabilistic structure leads to macroscopic phenomena unexpected by canonical statistical mechanics, which were verified by extensive simulations:
  \begin{itemize}
    \item FPUT family: A phase trajectory produces phase space configurations in concert with the Gibbs distribution of a fictitious ideal gas, even when the modal interaction is strong. Thermalization results from repulsion of different pieces of the trajectory (dubbed ``self-repulsion''), with each piece recording part of system's evolution history. The thermalization time depends on the maximal Lyapunov exponent and the initial condition in a way similar to the Ehrenfest time in quantum chaos \cite{Larkin69,Zaslavsky81,Izrailev90}.
    \item GPE family: A phase trajectory displays nonlinear wave localization in the mode space: The Rayleigh-Jeans (RJ) thermal equilibrium sets in the localization volume, and outside that volume the mode distribution decays exponentially. This suggests that, with classical wave effects being taken into account, classical physics suffices to cure the ultraviolet catastrophe.
  \end{itemize}

\begin{figure}[t]
      \centering
      \includegraphics[width=1\linewidth]{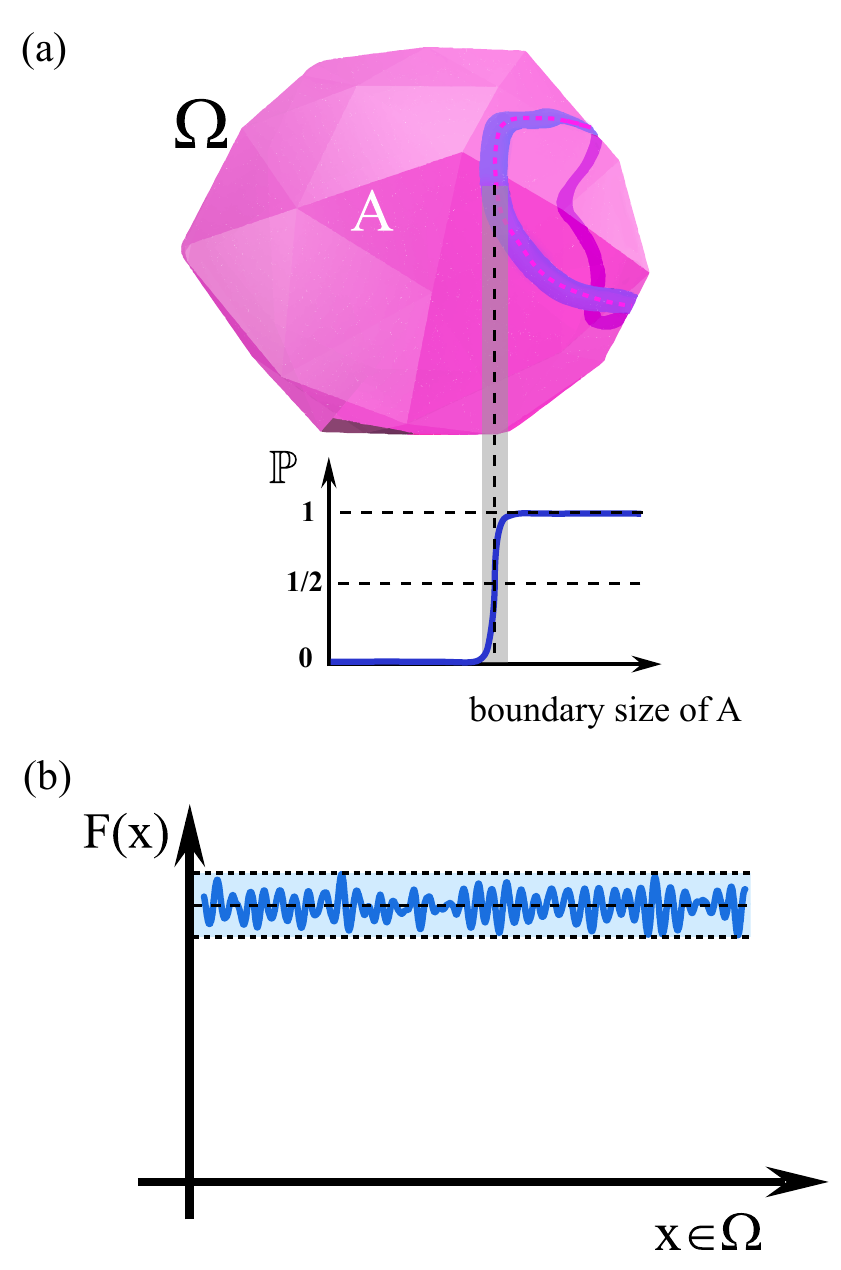}
      \caption{Illustration of concentration of measure. (a) The measure of a subset $A$ undergoes a sharp transition at $\mathbb{P}(A)=1/2$, where by slightly enlarging the boundary of $A$ the full measure is attained. Thus the full measure concentrates on an extremely small region. (b) This geometric phenomenon manifests in that any Lipschitz function, no matter how complicated it is, is essentially a constant, displaying small irregular oscillations.}
      \label{fig:1}
      \end{figure}

  Our theory (see SI \cite{SI} for details) starts from a geometric principle. Let $\Omega$ be a generic space equipped with metric $d$ and measure $\mathbb{P}$ normalized to unity. Like the celebrated isoperimetric theorem in the Euclidean plane, in $\Omega$ there are extremal sets minimizing the boundary among sets of same measure \cite{Ledoux01,Talagrand96}. In particular, for a family of $\Omega$ it has been proved (Fig.~\ref{fig:1}(a))
  that as long as a subset has measure $\geq\frac{1}{2}$, by slightly enlarging its boundary almost the full measure is attained, and as such the full measure concentrates on an extremely small region (e.g. a narrow strip around the equator on a high-dimensional sphere) \cite{Milman86}. Mathematically, this is encoded in the generalized isoperimetric inequality: For any subset $A$ of $\Omega$ with $\mathbb{P}(A)\geq\frac{1}{2}$, the set of points $x$ at distance to $A$, denoted $d(x,A)$, exceeding $\delta$ has measure
  \begin{equation}\label{eq:isoperimetry_inequality}
  \mathbb{P}(d(x,A)\geq\delta)\leq{\cal C}(\delta),
  \end{equation}
  and the concentration rate ${\cal C}$ decays rapidly with $\delta$. This geometric phenomenon, dubbed concentration of measure, can be ``visualized'' by a broad class of observables --- the functions $F(x)$ on $\Omega$ (Fig.~\ref{fig:1}(b)). The unique requirement is the Lipschitz continuity, i.e. the constant $L\equiv \sup_{x,x'}\frac{|F(x)-F(x')|}{d(x,x')}<\infty$. Indeed, thanks to Eq.~\eqref{eq:isoperimetry_inequality} no matter how complicated the functional form of $F$ is, $F$ is essentially a constant, and the probability for deviation $\geq\delta$ is bounded by $2{\cal C}(\delta/L)$. Below we show that the concentration-of-measure phenomenon is naturally embedded into the seemingly unrelated many-body Hamiltonian dynamics.

  We first consider a FPUT family \cite{Izrailev05,Gallavotti08} modeling one-dimensional anharmonic solids consisting of $N$ atoms ($N$ large but finite). Their Hamiltonian reads
  \begin{equation}\label{eq:Hamiltonian_FPUT}
    H=\sum_{j=0}^{N}{p_j^2\over 2}+\sum_{j=0}^{N-1}{(q_{j+1}-q_j)^2\over 2}+{u\over n} \sum_{j=0}^{N-1} ( q_{j+1} - q_j )^n,
  \end{equation}
  with $n = 4, 6, 8,...$. Here $p_j\,(q_j)$ is the momentum (position) of atom $j$ with $q_{0,N}=0$, and $u>0$ is the anharmonic interaction strength. $N$ phonon modes, each represented by canonical coordinates ($Q_k,P_k$) namely the
  Fourier transform of ($q_j,p_j$), interact, leading the power spectrum of $Q_k(t)$ to display a sharp peak of finite width $\tilde{\Gamma}_k$ \cite{He21,Lvov18}. The center frequency $\tilde{\omega}_k$ of the peak is shifted from the free phonon frequency $\omega_k= 2 \sin {\pi k\over 2N}$. Thus we can rewrite $H$ as
  \begin{eqnarray}\label{eq:2}
    \begin{split}
      H=\sum_k E_k+{\cal V}(Q),\quad E_k=(P_{k}^{2}+\tilde{\omega}_{k}^{2}Q_{k}^{2})/2,\\
      {\cal V}(Q)=V(Q)+\sum_k ({\omega}_{k}^{2}-\tilde{\omega}_{k}^{2})Q_{k}^{2}/2.\quad\quad
    \end{split}
  \end{eqnarray}
  Here $E_k$ is the energy carried by mode $k$, which differs from the free phonon mode in the frequency shift: $\omega_k\to \tilde{\omega}_{k}$. ${\cal V}$ accounts for the interaction among the modes of shifted frequency [$Q\equiv( Q_1,...,Q_N )$ and likewise for $P$ and others to appear later]. And $V$ is the Fourier transform of the last term in Eq.~\eqref{eq:Hamiltonian_FPUT}.

  For $N\gg 1$ this FPUT family is widely believed to be (almost) ergodic, yielding $\langle a\rangle_{t\rightarrow\infty}=\langle a\rangle_{\text{G}}$ for a dynamical observable $a(P,Q)$ \cite{Benettin94,Pettini08,Deutsch18,Flach17}. This trades the time average along a dynamical trajectory (the left-hand side) to the average with respect to the Gibbs distribution $\propto e^{-\beta H}$ (the right-hand side), where the inverse temperature $\beta$ is determined by $N \varepsilon = \langle H\rangle_{\text{G}}$, with $\varepsilon$ the specific energy. Starting from this well-known fact, now we take a major step forward, linking dynamics to geometry: Thanks to the harmonic term $Q_{k}^{2}$ in $E_k$, we cast the Gibbs average into
    \begin{eqnarray}
    \label{eq:10}
      \begin{split}
              \langle a\rangle_{\text{G}} = \int \prod_{k=1}^N \frac{dP_k}{\sqrt{2\pi /\beta}}\,e^{-\frac{\beta P_{k}^{2}}{2}}\left({\int d\mathbb{P} e^{-\beta {\cal V}}\,a\over \int d\mathbb{P} e^{-\beta {\cal V}}}\right),\quad\\
              d\mathbb{P} = \prod_{k=1}^N \frac{d\bar{Q}_k}{\sqrt{2\pi /\beta}}e^{-\frac{\beta\bar{Q}_{k}^{2}}{2}},\quad\quad\quad\quad\quad
      \end{split}
    \end{eqnarray}
  where the rescaling: $\bar{Q}_k=\tilde{\omega}_kQ_k$ was made for convenience. Automatically, the $N$-dimensional Euclidean space $\mathbb{R}^N\ni {\bar Q}$ is equipped with a Gaussian measure $\mathbb{P}$, which allows us to apply the general geometric principle above. On the one hand, the extreme set in the isoperimetric problem is now the half space, obtained from dividing $\mathbb{R}^N$ into two parts by a hyperplane \cite{SI}, and the measure concentration encoded by Eq.~\eqref{eq:isoperimetry_inequality} occurs. On the other hand, we can utilize the sharpness of the power spectrum and Hamiltonian equations to show the Lipschitz continuity of function ${\cal V}$ \cite{SI},
  \begin{equation}\label{eq:3}
    L=\sup_{\bar{Q},\bar{Q}'}{|{\cal V}(\bar{Q})-{\cal V}(\bar{Q}')|\over d(\bar{Q},\bar{Q}')}\sim \gamma \sqrt{N\varepsilon}<\infty,
  \end{equation}
  where $\gamma={\rm max}_k ({\tilde{\Gamma}_k/\tilde{\omega}_k})$. Thus we use ${\cal V}$ to visualize the concentration of $N$-dimensional Gaussian measure $\mathbb{P}$. Specifically, using Pisier's theorem \cite{SI} we obtain
      \begin{equation}\label{eq:4}
          \mathbb{P}(|{\cal V}-\mathbb{E}{\cal V}|\geq\delta)\leq 2e^{-2\delta^{2}/(\pi^2\sigma)}.
      \end{equation}
  It implies that ${\cal V}$ concentrates around the mean $\mathbb{E}{\cal V}$, with fluctuations controlled by the variance factor
  \begin{equation}\label{eq:5}
    \sigma={L^2/\beta}\sim N\gamma^2\varepsilon/\beta.
  \end{equation}
  These results are beyond the central limit theorem, in particular, because ${\cal V}$ is highly nonlinear in $\bar{Q}_1,...,\bar{Q}_N$.

  By Eq.~\eqref{eq:4} we can define a subset of $\mathbb{R}^N$, for which the deviation $|{\cal V} - \mathbb{E}{\cal V}|$ does not exceed some value $\propto \sqrt{\sigma} \sim \sqrt{N}$. The measure of its complement is very small. Thanks to this and $\frac{{\cal V}-\mathbb{E}{\cal V}}{\mathbb{E}{\cal V}} = O({1/\sqrt{N}})$ in that set, we have ${\int d\mathbb{P} e^{-\beta {\cal V}}\,a\over \int d\mathbb{P} e^{-\beta {\cal V}}} {=}\int d\mathbb{P} a$ for $N\gg 1$. So Eq.~\eqref{eq:10} simplifies to
  \begin{eqnarray}\label{eq:8}
        \langle a\rangle_{\text{G}}\stackrel{N\gg 1}{=}\int \prod_{k=1}^N\frac{dP_kdQ_k}{2\pi /(\beta\tilde{\omega}_{k})} e^{-\beta E_k}
        \,a(P,Q).
  \end{eqnarray}
  Setting $a=H$ gives
  \begin{equation}\label{eq:6}
        \varepsilon=1/\beta+\mathbb{E}{\cal V}/N,
  \end{equation}
  which determines $\beta$. Finally we obtain
  \begin{eqnarray}\label{eq:14}
   \langle a\rangle_{t\rightarrow\infty}
   \stackrel{N\gg 1}{=}
        \int \prod_{k=1}^N\frac{dP_kdQ_k}{2\pi /(\beta\tilde{\omega}_{k})} e^{-\beta E_k}\,a(P,Q).
  \end{eqnarray}
  This shows that at long time a phase trajectory behaves as a fictitious thermal ideal gas composed of modes with frequencies $\tilde{\omega}_k$, even when the anharmonic interaction is strong. That interaction enters into $\beta,\,\tilde{\omega}_k$ only. Applying Eq.~(\ref{eq:14}) to $a=P_k^2/2$, $\tilde{\omega}_k^2 Q_k^2/2$, we find the energy equipartition: $\langle P_k^2/2\rangle_{t\rightarrow\infty}=\langle \tilde{\omega}_k^2 Q_k^2/2\rangle_{t\rightarrow\infty}=1/(2\beta)$.

\begin{figure*}[htbp]
    \centering
    \includegraphics[width=1\linewidth]{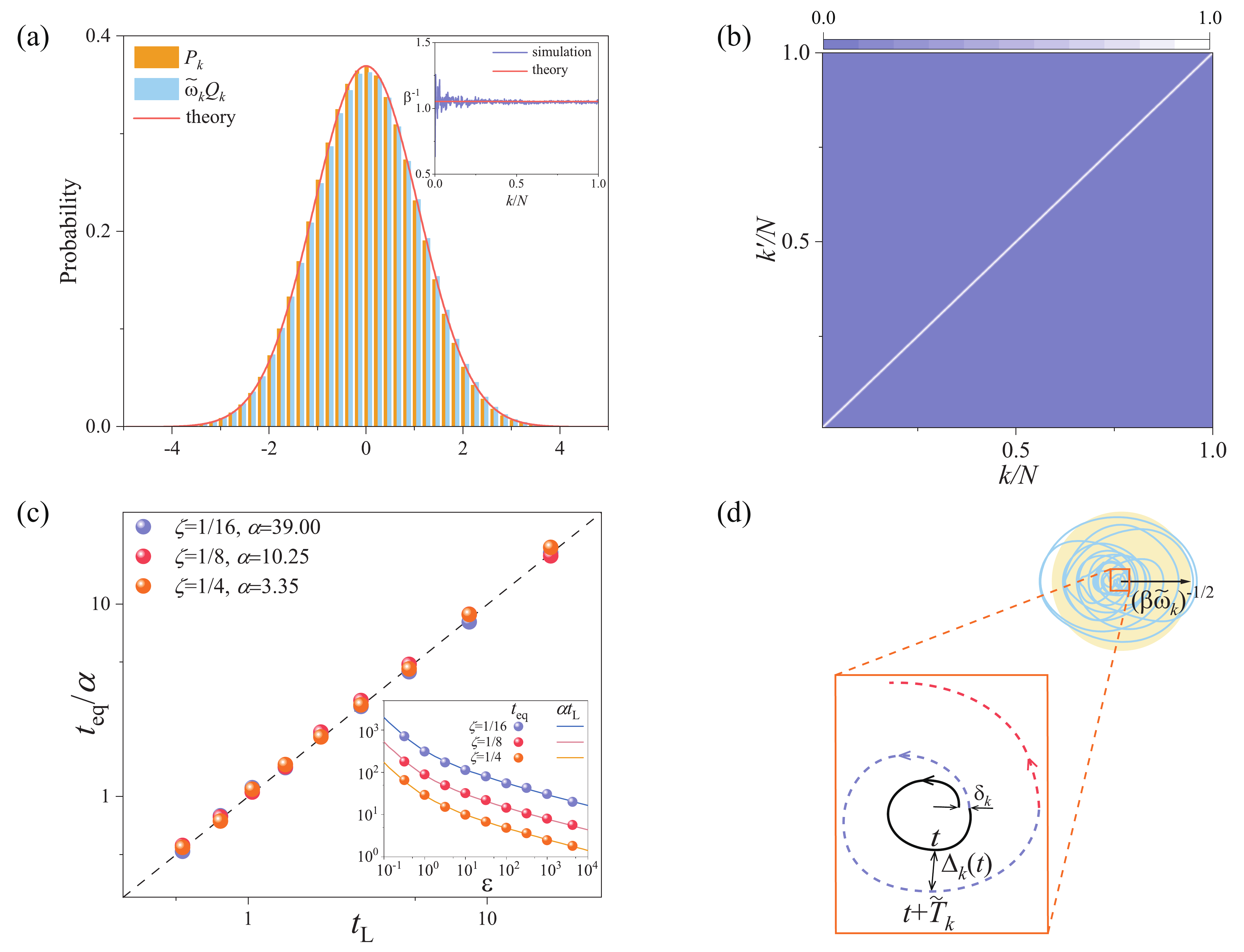}
    \caption{Many-body thermal equilibrium in FPUT dynamics. We simulate long-time FPUT dynamics and perform statistical analysis for a phase trajectory. (a) Simulated and theoretical distribution are in excellent agreement, with the comparison between simulated and theoretical temperature made in the inset. (b) The simulated correlation coefficient $r(E_k, E_{k'})$(see SI \cite{SI} for definition) confirms that different modes are uncorrelated, and thus the many-body equilibrium is described by the ideal-gas thermal distribution (\ref{eq:14}). $\varepsilon = 1$ and $\zeta = {1\over 8}$. (c) The relation (\ref{eq:1}) between the thermalization time and the maximal Lyapunov exponent (dashed line) is confirmed for a wide range of $\varepsilon$ and different $\zeta$ controlling the initial energy weight configuration. $\alpha$ in Eq.~(\ref{eq:1}) is seen to depend on $\zeta$ (inset). (d) Thermalization mechanism: A projective trajectory displays self-repulsion followed by irregular motion inside a region (grey), on which the thermal distribution concentrates. Lines in black, blue and red show deviations from periodic motion in the first three periods. For simulations $n=4$, $u=1$ and $N = 1025$.}
    \label{fig:2}
  \end{figure*}

  How much time does it take for the above thermal equilibrium to be approached? In particular, would it be the ergodic time, which increases exponentially with $N$? Canonical statistical mechanics cannot address this issue, since for an isolated system concepts such as ensemble evolution are invalid. Qualitatively, by Eq.~(\ref{eq:14}) at thermal equilibrium the projection of the trajectory onto the sub-phase space of mode $k$ visits $( P_k , Q_k )$ with Boltzmann distribution $\frac{\beta\tilde{\omega}_{k}}{2\pi}e^{-\beta E_k}$. Thus thermalization of mode $k$ requires the projective trajectory to deviate significantly from the periodic motion (with period $\tilde{T}_k$) and to irregularly explore a region of size ${1\over\sqrt{\beta\tilde{\omega}_k}}$ around the origin (Fig.~\ref{fig:2}(d)), where we make rescaling: ${P}_k \rightarrow {P_k\over \sqrt{\tilde{\omega}_{k}}}$, $Q_k \rightarrow \sqrt{\tilde{\omega}}_{k} Q_k$.

  The deviation from periodic motion differs from usual Lyapunov instability. To study it quantitatively we define (cf.~Fig.~\ref{fig:2}(d)) $\delta P_{k}(t)\equiv\tilde{\omega}_k^{-1/2}(P_{k}(t+\tilde{T}_{k})-P_{k}(t))$, $\delta Q_{k}(t)\equiv \tilde{\omega}_k^{1/2}(Q_{k}(t+\tilde{T}_{k})-Q_{k}(t))$, which measure the distance between an instantaneous state $(P_k(t),\,Q_k(t))$ in the projective trajectory and its future state with delay $\tilde{T}_k={2\pi/\tilde{\omega}_k}$. In SI \cite{SI} we find that
  \begin{eqnarray}\label{eq:12}
  {d\over dt}\left(\!\!
              \begin{array}{c}
                \delta Q_{k}^2 \\
                \delta P_{k}^2 \\
                \delta P_{k}\delta Q_{k} \\
              \end{array}
            \!\!\right)=
            \left(\!\!
              \begin{array}{ccc}
                0 & 0 & 2\tilde{\omega}_k \\
                {\kappa\over \tilde{\omega}_k^2}& 0 & -{2{\Omega}_0\over \tilde{\omega}_k} \\
                -{{\Omega}_0\over \tilde{\omega}_k} & \tilde{\omega}_k & 0 \\
              \end{array}
            \!\!\right)\!\!
            \left(\!\!
              \begin{array}{c}
                \delta Q_{k}^2 \\
                \delta P_{k}^2 \\
                \delta P_{k}\delta Q_{k} \\
              \end{array}
            \!\!\right),\quad
      \end{eqnarray}
  which gives
\begin{eqnarray}\label{eq:11}
  \Delta_k^2(t)\equiv\delta{Q}_{k}^2(t)+\delta {P}_{k}^2(t)=\delta^2_{k}e^{2t/t_{\text{L}}},~~\text{for}\,\,t\gtrsim t_{\text{L}}.\,\,\,\,
\end{eqnarray}
Here the parameters $\kappa,\Omega_0$ characterize the fluctuations of $\partial_{Q_k}^2{\cal V}$ and govern $t_{\text{L}}$, and $\delta^2_{k}$ depends on the initial phase coordinates. By Eq.~(\ref{eq:11}) the distance between a future and a past state --- with constant time delay $\tilde{T}_k$ --- increases exponentially with their central time. So self-repulsion results: The projective trajectory in the first period: $0\leq t\leq \tilde{T}_k$ repels that in the second: $\tilde{T}_k\leq t\leq 2\tilde{T}_k$, the latter repels that in the third, and so on (Fig.~\ref{fig:2}(d)). As such the periodic motion is inhibited. When $\Delta_k^2 = {1/(\beta\tilde{\omega}_k)}$ thermal equilibrium is established. This gives the thermalization time of mode $k$:
\begin{equation}\label{eq:9}
        t_{{\rm eq},k}=t_{\text{L}}\ln(\beta\tilde{\omega}_{k}\delta_{k}^{2})^{-{1\over 2}}.
    \end{equation}
We see that the interaction ${\cal V}$ is indispensable for approaching thermal equilibrium. Since $\Delta_k^2(t)$ is not the usual deviation of two phase trajectories, the meaning of $t_{\text{L}}^{-1}$ is not immediately clear. In SI \cite{SI} we show analytically that it is identical to system's maximal Lyapunov exponent. Thus $t_{\text{eq},k}$ is analogous to the Ehrenfest time, with Planck's quantum replaced by $(\beta\tilde{\omega}_{k}\delta_{k}^{2})^{1/2}$, and is dramatically smaller than the ergodic time.

If we fix a generic initial configuration of phases and energy weights ${E_k(0)\over \sum_{j=1}^{N}E_j(0)}$, because variations of $\varepsilon$ cannot change the logarithm in Eq.~(\ref{eq:9}) significantly, practically it can be treated as a numerical coefficient $\alpha$, depending on that configuration. So Eq.~(\ref{eq:9}) is simplified to a uniform thermalization time
\begin{eqnarray}
t_{\text{eq}}\approx \alpha t_{\text{L}},
\label{eq:1}
\end{eqnarray}
inversely proportional to the maximal Lyapunov exponent $t_{\text{L}}^{-1}$. Thus $t_{\text{eq}}$ and $t_{\text{L}}$ exhibit the same scaling behaviors, which will be further derived in SI~\cite{SI}.

We adopt the $10$th-order symplectic algorithm~\cite{Flach09} to simulate long-time FPUT dynamics. For studying thermalization, the initial energy weight is uniform for $0 < k \leq \zeta$($N - 1$) and zero for higher modes, and the initial phases are independent random variables. As shown in Fig.~\ref{fig:2}(a), simulations confirm that a trajectory produces configurations in a sub-phase space of mode $k$ in concert with the distribution: $\frac{\beta\tilde{\omega}_{k}}{2\pi}e^{-\frac{\beta}{2}(P_{k}^{2}+\tilde{\omega}_k^2{Q}_{k}^{2})}$. The value of $\beta$ is seen to be independent of $k$, consistent with the many-body distribution in Eq.~(\ref{eq:14}), and agrees with the theoretical value obtained from Eq.~(\ref{eq:6}). Simulations also show that at long time {energy of different modes are uncorrelated (Fig.~\ref{fig:2}(b)), and so are other observables~\cite{SI}. Thus at equilibrium different modes are uncorrelated, verifying that the system behaves as a thermal ideal gas. Finally, for different initial energy weight configurations, simulations verify that as $\varepsilon$ increases so that the transition from weak to strong nonlinearity occurs, Eq.~(\ref{eq:1}) holds always (Fig.~\ref{fig:2}(c)).}

We further generalize the developed theory to a family of one-dimensional GPE introduced in Refs.~\cite{Lewenstein00,Lewenstein01}, which suffers strong ergodicity breaking. The system is described by a wavefunction $\psi(z,t)$ (normalized to unity), satisfying
\begin{equation}
  i\partial_t \psi=(-\partial_z^2+g|\psi|^2)\psi.
\end{equation} 
We focus on strong nonlinearity, $g\ell\gg 1$, with $\ell$ being system's length. Importantly, the rigid boundary condition: $\psi(z,t)|_{z=0,\ell}=0$ is imposed,
so that the system is nonintegrable \cite{Lewenstein00,Lewenstein01}, whereas the system is integrable for periodic boundary condition \cite{Morris82}. With Fourier transform $\psi(z,t)=\sqrt{2\over \ell}\sum_{k}c_k(t)\sin{k \pi z\over \ell}$, this GPE is written as infinite-dimensional Hamiltonian equations of motion:
\begin{eqnarray}\label{eq:Hamiltonian_GPE}
\begin{split}
   {dc_k\over dt}={\partial H\over \partial (ic_k^*)}, \quad {d (ic_k^*)\over dt}=-{\partial H\over \partial c_k},\quad\quad\quad\\
   H=\sum_k \omega_k c_k^*c_k+{g\over 2\ell}\sum_{klmn}S(k,l,m,n)c^*_kc^*_lc_mc_n,
\end{split}
\end{eqnarray}
where $\omega_k=({\pi k/\ell})^2$ and $S$ gives the selection rule for the modal interaction (see SI \cite{SI} for its explicit form).
Initially we excite the lowest two modes.

Provided some {\it intrinsic} parameter $k_c$ exists, such that $\sum_{k> k_c}\langle|c_{k}^2|\rangle_{t\rightarrow\infty}\ll 1$, effectively the motion is restricted to a sub-phase space, the Euclidean space $\mathbb{R}^{2k_c}$ with coordinates $c_k^*,c_k$ ($k\leq k_c$). As shown below, such $k_c$ indeed exists, and $k_c\gg 1$ for large $g\ell$. Similar to FPUT dynamics, the power spectrum of $c_k(t)$ is sharply peaked at shifted frequency $\tilde{\omega}_k$. So we can write $H=\sum_k \tilde{\omega}_k c_k^*c_k+{\cal V}(c^*,c)$, where ${\cal V}$ is the modal interaction. In doing so, $\mathbb{R}^{2k_c}$ attains a Gaussian measure
\begin{equation}\label{eq:Gaussian_GPE}
  d\mathbb{P}=\prod_{k=1}^{k_c}\frac{dc^*_kdc_k}{2\pi /(\beta(\tilde{\omega}_{k}-\mu))}\,e^{-\beta (\tilde{\omega}_{k}-\mu) c^*_kc_k},
\end{equation}
where $\mu$ (the chemical potential) and $\beta$ are fixed by $(c^*(0),c(0))$.

So we can apply the geometric principle Eq.~\eqref{eq:isoperimetry_inequality} to $\mathbb{R}^{2k_c}$. In particular, it can be shown that ${\cal V}$ is Lipschitz continuous also, with $L$ governed by the peak width. Then similar to the derivation of Eq.~(\ref{eq:14}), for a dynamical observable $a(c^*,c)$ we obtain
\begin{eqnarray}\label{eq:17}
        \langle a\rangle_{t\rightarrow\infty}
        \stackrel{g\ell\gg 1}{=}
        \int\prod_{k=1}^{k_c}\frac{dc^*_kdc_k\,e^{-\beta (\tilde{\omega}_{k}-\mu) c^*_kc_k}}{2\pi /(\beta(\tilde{\omega}_{k}-\mu))}
        a(c^*,c).\quad
\end{eqnarray}
So, a trajectory $(c^*(t),c(t))$ displays behaviors in concert with a fictitious thermal ideal gas, composed of $k_c$ modes. For $a=|c_k|^2$ ($k\leq k_c$), Eq.~(\ref{eq:17}) gives the RJ distribution $n_{{\rm RJ},k}\equiv {1/(\beta(\tilde{\omega}_{k}-\mu))}$.

\begin{figure}[t]
    \centering
    \includegraphics[width=1\linewidth]{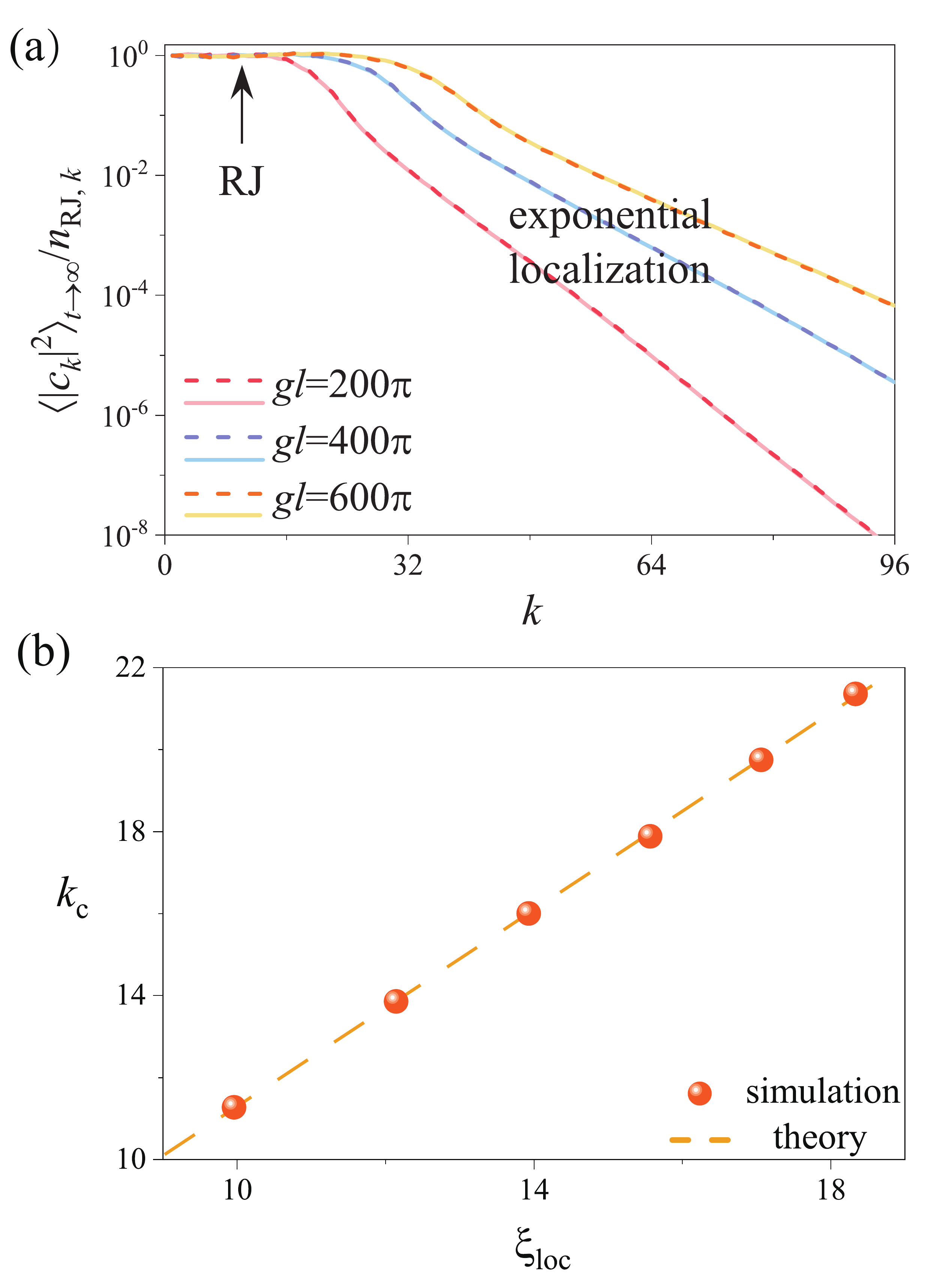}
    \caption{GPE escapes ultraviolet catastrophe. (a) Simulations confirm the emergence of the RJ equilibrium from a single trajectory for $k\leq k_c$, and exponential localization beyond $k_c$. For each $g\ell$, two sets of $(g,\ell)$ are used and the profiles overlap perfectly. (b) Verifications of the relation (\ref{eq:18}) between $\xi_{\rm loc}$ and $k_c$.}
    \label{fig:3}
\end{figure}

Why can an intrinsic $k_c$ exist and what happens beyond $k_c$ in the mode ($k$) space? To address these issues it suffices to consider long time, for which as shown above $c_k(t)=\sqrt{n_{{\rm RJ},k}}e^{i\tilde{\omega}_kt}$($k\leq k_c$). Substituting it into the nonlinear potential, defining $\theta\equiv \pi z/\ell$, and rescaling $t$ by $(\ell/\pi)^2$, we write the GPE as
\begin{equation}\label{eq:16}
i\partial_t \psi(\theta,t)=\left(\hat{h}_0+\hat{h}_1(\tilde{\omega}_1 t,...,\tilde{\omega}_{k_c} t)\right)\psi(\theta,t).
\end{equation}
The operator $\hat{h}_0$ is time independent, whereas $\hat{h}_1$ is quasiperiodic driving by $k_c$ incommensurate frequencies: $\tilde{\omega}_1,...,\tilde{\omega}_{k_c}$; see SI for their explicit form. Both depend on single parameter, $g\ell$. As we are interested in the time scale much larger than that of quasiperiodic oscillations, $\hat{h}_{1}$ behaves as an external driven potential. This kind of driven systems were studied long ago~\cite{Kravtsov03} and have received renewal interests recently~\cite{Chandran22}. Most importantly, they are found to display dynamical localization~\cite{Kravtsov03,Chandran22}, which is an analog of Anderson localization in the mode space. Thus we expect that the mode distribution decays exponentially for $k\gtrsim k_c$, i.e.
\begin{equation}\label{eq:distribution_GP}
        \langle |c_{k} |^{2}\rangle_t\xrightarrow{t\rightarrow\infty}
        \begin{cases}
            \sim e^{-k/\xi_{\text{loc}}},&k\gtrsim k_c,\\
            n_{{\rm RJ},k},&1\leq k\leq k_c,
        \end{cases}
\end{equation}
with $\xi_{\text{loc}}$ the localization length. For GPE is a nonlinear wave system, this localization phenomenon has classical wave and nonlinear nature, differing from many-body localization~\cite{Altshuler06} and its recent quantum dynamical analogs~\cite{Gupta22,Weld22,Ying23}. Since at $k\sim k_c$ the RJ equilibrium crosses over to localization, we have
\begin{equation}\label{eq:18}
\xi_{\text{loc}}\propto k_c.
\end{equation}
Both depend on $g,\ell$ through single parameter, $g\ell$. These results are confirmed by simulations (Fig.~\ref{fig:3}).

For $\mu=0$, Eq.~(\ref{eq:distribution_GP}) shows that the energy equipartition: $\tilde{\omega}_k\langle|c_k|^2\rangle_{t\rightarrow\infty} =1/\beta$ holds amongst modes $k\leq k_c$, while the modal energy decays as $e^{-k/\xi_{\text{loc}}}$ for $k\gtrsim k_c$. This explains the recent surprising numerical experiments on blackbody radiation \cite{Wang22}, and shows that, through nonlinear wave localization, an infinite-dimensional classical system can escape the ultraviolet catastrophe, contrary to the common belief for over $125$ years.

As our theory requires integrability breaking but not necessarily strong, it can be generalized to other many-body Hamiltonian systems, notably KAM systems. In this case the measure concentration would take place in some product space, similar to what occurs in quantum entanglement dynamics \cite{Tian24}. Moreover, the isoperimetric inequality (\ref{eq:isoperimetry_inequality}) would be replaced by Talagrand's convex distance inequality \cite{Talagrand96,Ledoux01}. We leave thorough investigations for future studies. The predicted nonlinear wave localization for GPE may be tested by cold atom experiments, especially because preparing a one-dimensional box to implement the rigid boundary condition, tuning the atomic interaction strength, and maintaining quantum coherence for long time are all well within the experimental reach nowadays \cite{Hadzibabic12,Chin10}. It may also find applications in self-cleaning effects in nonlinear optics \cite{Christodoulides22}.

\begin{acknowledgments}
  We thank Giancarlo Benettin, Jean-Claude Garreau, Xiaoquan Yu and Shizhong Zhang for inspiring discussions. Y.L. is the Yukawa Research Fellow supported by Yukawa Memorial Foundation. This work was supported by NSFC (Grants No.12475039, No. 12475043, No.12447101, No.11925507, No. 12075199 and No. 12247172), Guangdong Basic and Applied Basic Research Foundation (Grant No. 2025A1515010350). Numerical simulations were performed on TianHe-1 (A) at National Supercomputer Center in Tianjin.
\end{acknowledgments}

\end{document}